\documentclass[11pt,a4paper]{article}
\usepackage{jheppub}
\usepackage[T1]{fontenc}
\usepackage{amssymb}
\usepackage{mathtools}
\usepackage{stackengine}
\usepackage{scalerel}

\usepackage{hyperref}
\usepackage{epsfig}
\usepackage{amsmath,latexsym,amssymb}
\usepackage{graphicx}
\usepackage{braket}
\usepackage{pdfsync}

\usepackage{framed}
\usepackage{mathtools}

\usepackage{jheppub}
\usepackage[T1]{fontenc}
\usepackage{amssymb}
\usepackage{mathtools}
\usepackage{amsmath}
\usepackage{bm}
\usepackage{stackengine}

\usepackage{scalerel}
\usepackage{eufrak}
\usepackage{framed}

\usepackage{mathtools}

\usepackage{pstricks}

\usepackage{amsmath}
\usepackage{bbold}
\usepackage{bm}
\usepackage{latexsym}
\usepackage{braket}
\usepackage{slashed}
\usepackage{graphicx,booktabs,multirow}
\usepackage{eufrak}
\numberwithin{equation}{section}

\usepackage{color}
\usepackage{pstricks}
\usepackage{amssymb}

\makeatletter
\newcommand{\doublewidetilde}[1]{{%
  \mathpalette\double@widetilde{#1}%
}}
\newcommand{\double@widetilde}[2]{%
  \sbox\z@{$\m@th#1\widetilde{#2}$}%
  \ht\z@=.9\ht\z@
  \widetilde{\box\z@}%
}
\makeatother

\usepackage{hyperref}
\usepackage{slashed}
\usepackage{epsfig}
\usepackage{amsmath,latexsym,amssymb}
\usepackage{graphicx}
\usepackage[latin1]{inputenc}
\usepackage{braket}
\usepackage{pdfsync}

\usepackage{framed}

\usepackage{mathtools}

\def\be{\begin{equation}}
\def\ee{\end{equation}}
\def\ba{\begin{eqnarray}}
\def\ea{\end{eqnarray}}

\newcommand{\bz}{\bar{z}}

\newcommand{\bh}{\bar{h}}

\newcommand{\bx}{\bar{x}}

\newcommand{\zbar}{\bar{z}}

\def\IR{{\bf R}}
\def\IC{{\bf C}}

\def\IZ{{\bf Z}}

\def\IZ{ {\bf Z}}\def\IC{{\bf C}}\def\IR{ {\bf R}}

\def\D{\Delta}

\def\p{\pi}

\def\cG{{\cal{G} }}

\newcommand{\comment}[1]{}

\def\fc#1#2{{\frac{#1}{#2}}}
\newcommand{\req}[1]{(\ref{#1})}

\def\Ic {{\cal I}}

\def\Ic{{\cal I}}
\def\h{\fc{1}{2}}

\def\p{\partial}

\newcommand{\eea}{\end{eqnarray}}
\def\lf{\left}
\def\ri{\right}

\author{
Wei Fan${}^{1}$, Angelos Fotopoulos${}^{2}$, Stephan Stieberger${}^{3}$,
Tomasz R.\ Taylor${}^{2}$,\, Bin Zhu${}^2$\\[0.5cm] $^1$\it
Department of Physics, School of Science, Jiangsu University of Science and Technology,
Zhenjiang, 212003, China\\[0.2cm]
 $^2${\it Department of Physics \\
  Northeastern University, Boston, MA 02115, USA}\\[0.2cm]
$^3${\it Max--Planck--Institut f\"{u}r Physik,	Werner--Heisenberg--Institut, \\80805 M\"unchen, Germany}}
\emailAdd{fanwei@just.edu.cn}
\emailAdd{a.fotopoulos@northeastern.edu}
\emailAdd{stephan.stieberger@mpp.mpg.de}
\emailAdd{taylor@neu.edu}
\emailAdd{zhu.bi@northeastern.edu}

\title{\boldmath \centerline{Celestial Yang-Mills Amplitudes and } \hskip 4 cm $D\!=\! 4$ Conformal Blocks \unboldmath}

\abstract{We discuss the properties of recently constructed ``single-valued'' celestial four-gluon amplitudes. We show that the
amplitude factorizes into the ``current'' part and the ``scalar'' part. The current factor is given by the group-dependent part of the Wess-Zumino-Witten correlator of four holomorphic currents with a non-vanishing level of Ka\v{c}-Moody algebra. The scalar factor can be expressed in terms of a complex integral of the Koba-Nielsen form, similar to the integrals describing four-point correlators in Coulomb gas models and, more generally, in the infinite central charge limit of Liouville theory. The scalar part can be also obtained by a dimensional reduction of a single $D=4$ conformal block and the shadow block from Minkowski space to the celestial sphere.}


\begin{document}
\maketitle
\section{Introduction}
Classical Yang-Mills (YM) theory is invariant under $O(4,2)$ conformal transformations. In quantum theory, at the tree level, gluon scattering amplitudes are conformally invariant, but this invariance is broken by loop corrections. Furthermore, there is an overwhelming evidence for the existence of a mass gap in YM theory. Hence studying conformal properties of gluon amplitudes should lead to a better understanding of conformal symmetry breaking in gauge theories.

There had not been much attention given to the conformal properties of YM scattering amplitudes until Witten formulated perturbative YM as a string theory in conformally-invariant twistor space in 2003 \cite{Witten:2003nn}. More recently, two of us \cite{Stieberger:2018onx} examined the symmetries of celestial amplitudes \cite{Pasterski:2017ylz,Pasterski:2021rjz} obtained by applying Mellin transforms to the conventional (momentum space) amplitudes that convert them into the correlators of a putative two-dimensional conformal theory on the celestial sphere (CCFT). In CCFT, the dimensions of primary field operators associated to gluons are $\Delta_n=1+i\lambda_n$, with real $\lambda_n$ \cite{Pasterski1705}. The dilatation generator \cite{Stieberger:2018onx} acts on the celestial amplitudes as
\be {\cal D}=-i\sum_n(\Delta_n-1)=\sum_n\lambda_n\ , \ee
therefore scale-invariant tree-level gluon amplitudes are non-zero only if $\sum_n\lambda_n=0$.
This constraint is enforced in all celestial gluon amplitudes by the universal delta-function factor
 $\delta(\sum_n\lambda_n)$. Furthermore, the invariance under special conformal transformations implies a set of second order differential equations for celestial amplitudes.

In celestial holography, the directions of four-dimensional momenta are mapped into the positions of primary field operators. As a result, the momentum conservation law imposes constraints \cite{Mizera:2022sln} on these positions, what, from the point of view of two-dimensional CFT, is an unwanted feature of the correlation functions. In a recent work \cite{Fan:2022vbz}, we circumvented these constraints by constructing the ``single-valued'' CCFT correlation functions. They are obtained from the Mellin transforms of the scattering amplitudes evaluated in the presence of a dilaton background with the source located on the celestial sphere. While this background violates translational symmetry, it preserves scale invariance. In this work, we discuss the properties of ``single-valued'' CCFT correlators. We find new representations of the celestial four-gluon correlator and discuss its relation to
the limits of $D=4$ conformal blocks at null infinity.

\section{Integral representation of the single-valued four-gluon amplitude}
\def\Dc{{\cal D}}
The four-gluon celestial MHV amplitude is expressed in terms of two partial amplitudes, $M(z_i,\bz_i)$ and $\widetilde M(z_i,\bz_i)$ in the following way
\cite{Fan:2022vbz}:
\begin{align}
\Big\langle\phi_{\D_1,-}^{a_1,-\epsilon}(z_{1},\zbar_{1}
)\,
\phi_{\D_2,-}^{a_2,-\epsilon}&(z_2,\bz_2)
\,\phi_{\D_3,+}^{a_3,+\epsilon}
(z_3,\bar z_3)
\,\phi_{\D_4,+}^{a_4,+\epsilon}(z_4,\bz_4)\Big\rangle= \nonumber\\[1mm] &~~=f^{a_1a_2b}f^{a_3a_4b}M(z_i,\bz_i)
+f^{a_1a_3b}f^{a_2a_4b} \widetilde M(z_i,\bz_i) .\label{partials}
\end{align}
Here, the subscripts $\Delta_i,J_i=\pm 1$ refer to the dimensions and spins (gluon helicities) of the primary fields, respectively, group indices are labeled by $a_i$, while the superscripts $\pm\epsilon$ indicate incoming
$(-\epsilon)$ and outgoing $(+\epsilon)$ particles.
Taking advantage of conformal symmetry, one defines
\begin{align}
G(x,\bar{x}) &= \lim_{z_1,\bar{z}_1\rightarrow \infty} z_1^{2 h_1} \bar{z}_1^{2\bar{h}_1} M (z_1; z_2=1; z_3=x; z_4=0) \, , \\
\widetilde G(x,\bar{x}) &= \lim_{z_1,\bar{z}_1\rightarrow \infty} z_1^{2 h_1} \bar{z}_1^{2\bar{h}_1} \widetilde{M}(z_1; z_2=1; z_3=x; z_4=0)\, ,
\end{align}
with the conformal weights $h_1=(\Delta_1+J_1)/2, \bh_1=(\Delta_1-J_1)/2$. The complex variable $x$
is identified with the complex cross ratio
\be x= \frac{z_{12}z_{34}}{z_{13}z_{24}}\ .\label{xdef}\ee
{}For the future use, it is convenient to define another cross ratio,
\be
z=\fc{z_{12}z_{43}}{z_{14}z_{23}}=\fc{x}{x-1}\label{zdef}
\ee
and real conformal invariants
\be u=\lf|\fc{x}{x-1}\ri|^2=|z|^2\  ,\qquad
v=\fc{1}{|1-x|^2}=|1-z|^2\
.\label{variables1}
\ee
Note that upon interchanging $3\leftrightarrow 4$, $x\leftrightarrow z$.

The partial amplitudes can be written as
\be
M(z_i,\bz_i)= \Pi(z_i,\bz_i)\,G(x,\bar{x})\ ,\qquad \widetilde M(z_i,\bz_i)= \Pi(z_i,\bz_i)\, \widetilde G(x,\bar{x})\ ,  \label{mmms}\ee
where
\begin{align}\Pi(z_i,\bz_i)&=
~z_{12}^{ ~ -h_1-h_2+h_3+h_4} z_{13}^{ ~-2h_3} z_{14}^{h_2+h_3-h_1-h_4}z_{24}^{h_1-h_2-h_3-h_4}  \nonumber\\
&\quad~\times \bar{z}_{12}^{ ~ -\bar{h}_1-\bar{h}_2+\bar{h}_3+\bar{h}_4} \bar{z}_{13}^{ ~-2\bar{h}_3} \bar{z}_{14}^{\bar{h}_2+\bar{h}_3-\bar{h}_1-\bar{h}_4}
\bar{z}_{24}^{\bar{h}_1-\bar{h}_2-\bar{h}_3-\bar{h}_4} .  \label{mmmss}
\end{align}
With the dimensions parameterized as $\Delta_n=1+i\lambda_n$, $\lambda_n\in \IR$, the conformal prefactor (\ref{mmmss}) becomes
\be
\Pi(z_i,\bz_i)= z_{12}^{~ -i \lambda_1- i\lambda_2+2} z_{13}^{~ -2-i\lambda_3}  z_{14}^{~ -i\lambda_1-i\lambda_4} z_{24}^{i\lambda_1-2}\bar{z}_{12}^{~ -i \lambda_1- i\lambda_2-2} \bar{z}_{13}^{~-i\lambda_3} \bar{z}_{14}^{~ -i\lambda_1-i\lambda_4}\bar{z}_{24}^{i\lambda_1}\ .
\ee
There are three ways of constructing the ``single-valued'' four-gluon correlators, all of them leading to the same result.

In Refs.\ \cite{Fan:2021isc,Fan:2021pbp}, we started from the celestial amplitude of Pasterski, Shao and Strominger \cite{Pasterski:2017ylz} and performed the shadow transformation on one of gluons. The shadow transform relaxes the kinematic constraints due to momentum conservation, but it leads to a correlator that is a multi-valued function of complex coordinates. It can be used, however, to construct a single-valued correlator by applying the same procedure as in minimal models \cite{DiF}, by adding a contribution of additional conformal blocks. At the end, we inverted the shadow transformation and obtained a well-defined ``unshadowed'' four-gluon correlator.
In \cite{Fan:2021pbp}, this procedure was performed explicitly only in the ``soft'' limit $\Delta\to 1$ of the shadowed gluon, but it can be easily extended to arbitrary $\Delta$.

More recently \cite{Fan:2022vbz}, we constructed this correlator by solving the differential equations written by Banerjee and Ghosh (BG) in Ref.\ \cite{Banerjee:2020vnt} and further developed in Ref.\cite{Hu:2021lrx}. BG equations follow from the consistency of soft theorems with the operator product expansions of CCFT. They
are similar to the equations describing the decoupling of null states in minimal models. In Ref.\ \cite{Fan:2022vbz}, we also showed that the same correlator describes four-gluon scattering in the presence of a dilaton background. This (third) representation of the single-valued celestial gluon amplitude is given by the following integral:
\begin{align}
~~G (x,&\bar{x})=\delta(\lambda_1+\lambda_2+\lambda_3+\lambda_4)\ \frac{1}{x(1-x)} \label{eq:G1_IntRep}\\
&\times \Bigg\{ \int_0^1 dt\, B(-i\lambda_1,-i\lambda_2)(1+i\lambda_1+i\lambda_2)\, t^{i\lambda_3-1}(1-t)^{i\lambda_4-1} \nonumber\\
&\times \big(t(1-x)(1-\bar{x})+1-t\big)^{i\lambda_1}  \, _2F_1\left({-i\lambda_1,-i\lambda_2 \atop -1-i\lambda_1-i\lambda_2}; \frac{t(1-t) \, x\, \bar{x}}{t (1-x)(1-\bar{x}) +1-t } \right) \nonumber\\
& +\int_0^1 dt \, B(2+i\lambda_1,2+i\lambda_2)(3+i\lambda_1+i\lambda_2)\, t^{1-i\lambda_4}(1-t)^{1-i\lambda_3} (x\bar{x})^{2+i\lambda_1+i\lambda_2}\nonumber\\
&\times\big(t(1-x)(1-\bar{x})+1-t\big)^{-2-i\lambda_2}\, _2F_1\left( {2+i\lambda_1, 2+i\lambda_2 \atop 3+i\lambda_1+i\lambda_2}; \frac{t(1-t) \, x\, \bar{x}}{t (1-x)(1-\bar{x}) +1-t }\right) \Bigg\} \, , \nonumber
\end{align}
where $_2F_1$ denotes Gau\ss\ hypergeometric function.
Similarly,
\be \widetilde{G}(x,\bx)=-xG(x,\bx)\ee

By a series of algebraic manipulations and changes of integration variables described in Appendix A, Eq.(\ref{eq:G1_IntRep}) can be recast as
\begin{align}
G (x,&\bar{x})=\delta(\lambda_1+\lambda_2+\lambda_3+\lambda_4)\ \frac{|1-x|^{2i\lambda_1}}{x(1-x)}\ \label{DuffinLook}\\
&\times\Dc_z\;\Bigg\{
B(-1+i\lambda_4,i\lambda_3)\; B(-i\lambda_2,-1-i\lambda_1) \nonumber\\
&~~~~~~~~~~\times {}_2F_1\left( {-1+i\lambda_4, -1-i\lambda_1\atop-1+i\lambda_3+i\lambda_4}; z\right)\; _2F_1\left( {-1+i\lambda_4, -1-i\lambda_1\atop-1+i\lambda_3+i\lambda_4}; \bar z\right)\nonumber\\[3mm]
&-B(1-i\lambda_3,2-i\lambda_4)\; B(2+i\lambda_1,1+i\lambda_2) \nonumber\\[1mm]
&~~~\lf.\times \,|z|^{2(2+i\lambda_1+i\lambda_2)}\; _2F_1\left( {1-i\lambda_3, 1+i\lambda_2\atop 3+i\lambda_1+i\lambda_2}; z\right)\; _2F_1\left( {1-i\lambda_3, 1+i\lambda_2\atop 3+i\lambda_1+i\lambda_2}; \bar z\right)
\Bigg\}\right|_{z=\fc{x}{x-1}},\nonumber
\end{align}
with the differential operator:
\be
\Dc_z=\fc{1}{z-\bar z}\ \lf(z\;\p_z-\bar z\;\p_{\bar z}\ri)= -\fc{\p}{\p v}\ . \label{opdcz}
\ee
The expression in the curly bracket of \req{DuffinLook} can be related to the complex integrals considered by Dotsenko and Fateev \cite{Dotsenko:1984ad, Dotsenko:1984nm} in the context of Coulomb gas models. To that end, we define
\begin{equation}\label{complexI}
\Ic(z,\bar z)=\int d^2w\ w^{\hat a+a}\ \bar w^{\hat a+\bar a}\ (w-1)^{\hat b+b}\ (\bar w-1)^{\hat b+\bar b}\ (w-z)^{\hat c+c}\ (\bar w-\bar z)^{\hat c+\bar c}\ ,
\end{equation}
with the parameters $\hat a,\hat b,\hat c\notin\IZ$ and $a,b,c,\bar a,\bar b,\bar c\in \IZ$.
The integral \req{complexI} can be evaluated through analytic continuation, by disentangling the holomorphic and antiholomorphic parts \cite{Dotsenko}. This procedure, which is also known as the Kawai-Lewellen-Tye (KLT) method, yields
\begin{align}
\Ic(z,\bar z)&=\frac{s(\hat b)s(\hat a+\hat b+\hat c)}{s(\hat a+\hat c)}\ I_1(\hat a+a,\hat b+b,\hat c+c;z)\ I_1(\hat a+\bar a,\hat b+\bar b,\hat c+\bar c;\bar z)\nonumber\\
&+\ (-1)^{b+\bar b+c+\bar c}\ \frac{s(\hat a)s(\hat c)}{s(\hat a+\hat c)}\  I_2(\hat a+a,\hat b+b,\hat c+c;z)\ I_2(\hat a+\bar a,\hat b+\bar b,\hat c+\bar c;\bar z)\ ,
\end{align}
with $s(\alpha):=\sin(\pi \alpha)$ and
\begin{align}
 I_1(a,b,c;z)&=\int_1^\infty dw\ w^a\ (w-1)^b\ (w-z)^c\nonumber\\
 &=B(-a-b-c-1,b+1)\ {}_2F_1\left({-c,-a-b-c-1\atop -a-c};z\right)\ ,\label{Int1}\\
I_2(a,b,c;z)&=\int_0^z dw\ w^a\ (1-w)^b\ (z-w)^c\nonumber\\
&=z^{1+a+c}\ B(a+1,c+1)\ {}_2F_1\left({-b,1+a\atop a+c+2};z\right)\ .\label{Int2}
\end{align}
By comparing the above equations with (\ref{DuffinLook}), we obtain
\begin{align}
G (x,\bar{x})&=\fc{1}{\pi}\;\delta(\lambda_1+\lambda_2+\lambda_3+\lambda_4)\
\frac{|1-x|^{2i\lambda_1}}{x(1-x)}\nonumber \\
&\times \lf.(2+i\lambda_1+i\lambda_2)\;B(2+i\lambda_1,1+i\lambda_2)\; B(i\lambda_3,-1+i\lambda_4)\ \Dc_z\;\Ic(z,\bar z)\right|_{z=\fc{x}{x-1}}\ ,\label{FinalCoul}
\end{align}
with the assignments:
\begin{equation}\label{abc}
\Ic(z,\bar z)\ \ \left\{\begin{array}{ccccc}
\hat a=-i\lambda_3\ &,&\ a=\bar a=0\ ,\\
\hat b=-i\lambda_2\ &,&\ b=\bar b=-1\ ,\\
\hat c=-i\lambda_4\ &,&\ c=\bar c=1\ .\\
\end{array}\right.
\end{equation}
To be more explicit, we can replace the cross ratio $z$ by $x$ and express \req{FinalCoul} in terms of
\be
\lf.\Ic(z,\bar z)\right|_{z=\fc{x}{x-1}}=|1-x|^{-2(i\lambda_1+1)}\;\int d^2w\ |w|^{-2i\lambda_4+2}\ |w-1|^{-2i\lambda_2-2}\ |w-x|^{-2i\lambda_3}\ ,
\ee
and
\be
\Dc_z=\fc{|1-x|^2}{x-\bar x}\;\lf\{x(1-x)\;\p_x-\bar x(1-\bar x)\;\p_{\bar x}\ri\}\ . \label{opdczx}
\ee
The fact that the four-gluon celestial amplitude can be expressed in terms of the integrals encountered in Coulomb gas models points to a deeper relation between CCFT and the Coulomb gas models and more generally, to a relation between CCFT and the complexified Liouville theory.

\section{Relation to Aomoto--Gelfand hypergeometric functions}
In this section, we wish to discuss an interesting relation of the four-gluon celestial amplitude to the Aomoto--Gelfand hypergeometric functions. This relation remains somehow outside the main scope of this paper, therefore it can be skipped at the first reading.

The Aomoto--Gelfand hypergeometric functions are constructed in the framework of Grassmannians \cite{Aomoto}.
Among some well-known examples, there are four hypergeometric functions of two variables, introduced by Appell. Here, we focus on
Appell's $F_4$, which is given by the following integral \cite{Aomoto}
\begin{align}
F_4\lf[\alpha,\beta,\gamma,\gamma';u,v\ri]&=\fc{\Gamma(1-\alpha)}{\Gamma(1-\gamma)\Gamma(1-\gamma')\Gamma(\gamma+\gamma'-\alpha-1)}\nonumber\\
&\times\int_{\Delta_2}dx_1\;dx_2\;x_1^{-\gamma}\;x_2^{-\gamma'}\;(1-x_1-x_2)^{\gamma+\gamma'-\alpha-2}\;\lf(1-\fc{u}{x_1}-\fc{v}{x_2}\ri)^{-\beta}
\end{align}
over the two-dimensional simplex:
\be
\Delta_2=\{(x_1,x_2)\in\IR^2\ |\ |x_1|+|x_2|\leq 1\}\ .
\ee

Appell's $F_4$ has the following power series expansion  around $u,v=0$:
\be\label{powerF4}
F_4\lf[\alpha,\beta,\gamma,\gamma';u,v\ri]=\sum_{m,n=0}^\infty\fc{(\alpha)_{m+n}\;
(\beta)_{m+n}}{(\gamma)_m(\gamma')_n}\ \fc{u^mv^n}{m!n!}\ ,\ u,v\in\IC\ ,\ |u|^{1/2}+|v|^{1/2}<1\ .
\ee
The above series solves a system of two partial (hypergeometric) differential equations called the $F_4$ equations. The general solution of the $F_4$ system can be written as a linear combination of four independent functions \cite{Goto}:
\begin{align}
f_1&=F_4\lf[\alpha,\beta,\gamma,\gamma';u,v\ri]\ ,\nonumber\\
f_2&=v^{1-\gamma'}\;F_4\lf[\alpha+1-\gamma',\beta+1-\gamma',\gamma,2-\gamma';u,v\ri]\ ,\nonumber\\
f_3&=u^{1-\gamma}\;F_4\lf[\alpha+1-\gamma,\beta+1-\gamma,2-\gamma,\gamma';u,v\ri]\ ,\nonumber\\
f_4&=u^{1-\gamma}\;v^{1-\gamma'}\;F_4\lf[\alpha+2-\gamma-\gamma',\beta+2-\gamma-\gamma',2-\gamma,2-\gamma';u,v\ri]\ .\label{basisF4}
\end{align}

The relation between the four-gluon celestial amplitude and the $F_4$ system can be obtained by
starting from \req{DuffinLook} and applying various hypergeometric function relations. The variables $u$ and $v$ are related to the cross ratios $x$ and $z$ according to Eq.(\ref{variables1}). The details of the derivation are given in Appendix B. At the end, the amplitude can be  expressed in the basis \req{basisF4} as
\begin{align}
G(x,\bar x)&=-\delta(\lambda_1+\lambda_2+\lambda_3+\lambda_4)\ \fc{1}{x(1-x)}
\ v^\beta\ \Bigg\{\fc{\Gamma(1-\gamma')\Gamma(\alpha)\Gamma(\beta)}{\Gamma(\gamma)}\;f_1\nonumber\\
&+\fc{\Gamma(\gamma'-1)\Gamma(\alpha+1-\gamma')\Gamma(\beta+1-\gamma')}{\Gamma(\gamma)}\;f_2
+\fc{\Gamma(1-\gamma')\Gamma(\alpha+1-\gamma)\Gamma(\beta+1-\gamma)}{\Gamma(2-\gamma)}\;f_3\nonumber\\
&+\fc{\Gamma(\gamma'-1)\Gamma(\alpha+2-\gamma-\gamma')\Gamma(\beta+2-\gamma-\gamma')}{\Gamma(2-\gamma)}\;f_4\Bigg\}\ ,\label{ExplicitF4}
\end{align}
with
\begin{align}
\alpha=i\lambda_4\ \ \ &,\ \ \ \beta=-i\lambda_1\ ,\nonumber\\
\gamma=-1+i\lambda_3+i\lambda_4\ \ \ &,\ \ \ \gamma'=1+i\lambda_2+i\lambda_4\ , \end{align}
satisfying \be \alpha+\beta=\gamma+\gamma'\ .
\ee
The integral representation (\ref{FinalCoul}), written as in Eq.(\ref{ExplicitF4}) and expanded by using
the series (\ref{powerF4}),
agrees with the series expansion solution of BG equations written explicitly in Section 2.1 of
Ref.\cite{Fan:2022vbz}.

\section{Relation to $D=4$ conformal blocks}
In this section, we establish a connection between celestial Yang-Mills amplitudes and conformal blocks in $D=4$ CFT. The relevant blocks appear in the four-point correlation functions of scalar (spin zero) $D=4$ fields with dimensions $d_n=i\lambda_n$. They depend on the spacetime coordinates $x_n^\mu$ of scalar fields. Due to $O(4,2)$ invariance, apart from a standard conformal prefactor, they depend on only two cross ratios,
\begin{equation}
U = \frac{x_{12}^{2}x_{34}^2}{x_{13}^2 x_{24}^2}\, , \quad V = \frac{x_{14}^2 x_{23}^2}{x_{13}^2 x_{24}^2} \, .
\end{equation}
It is often convenient to use another set of variables, $X$ and $\bar X$, defined by
\begin{equation}
U= X\bar X\,  , \quad V = (1-X) (1-\bar{X}) \, .\label{UV}
\end{equation}

The four-scalar correlator can be decomposed into conformal blocks associated to various representations of the conformal group.
A single conformal block associated to a scalar field of dimension $\Delta$ contributes \cite{Dolan:2000ut,Dolan:2003hv,Simmons-Duffin:2012juh}
\begin{equation}
W_\Delta(x_n) = \frac{1}{(x_{12}^2)^{\frac{d_1+d_2}{2}}(x_{34}^2)^{\frac{d_3+d_4}{2}}}\Big( \frac{x_{14}^2}{x_{13}^2}\Big)^{\frac{d_{34}}{2}}\Big( \frac{x_{24}^2}{x_{14}^2}\Big)^{\frac{d_{12}}{2}} g_\Delta(U,V) \, , \label{Duffinbls}
\end{equation}
where $d_{ij}=d_i-d_j$ and, with the cross ratios parameterized as in (\ref{UV}),
\begin{align}
g_\Delta(X,\bar X)&=\fc{1}{X-\bar X}\; |X|^\Delta \label{4dgblexp}\\
&\times\lf\{X\, _2F_1\left( {\h(\Delta-d_{12}),\h(\Delta+d_{34})\atop \Delta}; X\right)
{}_2F_1\left( {\h(\Delta-d_{12})-1,\h(\Delta+d_{34})-1\atop \Delta-2}; \bar X \right)
\ri.\nonumber\\
&\lf.-\bar X\, _2F_1\left( {\h(\Delta-\Delta_{12})-1,\h(\Delta+d_{34})-1\atop \Delta-2}; X\right)
{}_2F_1\left( {\h(\Delta-d_{12}),\h(\Delta+d_{34})\atop \Delta}; \bar X \right)\right\}.\nonumber
\end{align}

We are interested in the ``celestial limit'' of the conformal block
(\ref{Duffinbls}), leading from $D=4$ Minkowski space to $D=2$ celestial sphere. To that end, we use Milne coordinates \cite{deBoer:2003vf,Cheung:2016iub}, with
\begin{equation}
x^{\mu}=e^{\tau}\Big(\frac{k^{\mu}}{\rho}+\rho q^{\mu}\Big) \, ,
\end{equation}
where $k^{\mu}$ and $q^{\mu}$ are null vectors
\begin{equation}
k^{\mu}=\frac{1}{2}(1+z\bar{z},z+\bar{z},-iz+i\bar{z},-1+z\bar{z}) \quad \text{and} \quad q^{\mu} = \frac{1}{2}(1,0,0,1) \, .
\end{equation}
These coordinates describe a foliation of $D=4$ Minkowski space, divided into the Milne  and Rindler regions. Milne coordinates are used in the Milne region, while Rindler coordinates are used in the Rindler region \cite{deBoer:2003vf,Cheung:2016iub}.
In the limit of $\rho\rightarrow 0$,
\begin{equation}
x^{\mu} \to \frac{e^{\tau}k^{\mu}}{\rho}\ ,\qquad
x_{ij}^2 \to \frac{e^{\tau_i+\tau_j}}{\rho^2} \, z_{ij} \bar{z}_{ij} \, .
\end{equation}
Furthermore, we choose common $\tau$ for all four $x^{\mu}$, and  take $\tau\rightarrow +\infty$. This limit sends the spacetime coordinates of scalar fields to null infinity.\footnote{A similar limit can be taken in Rindler coordinates. Such a limit can be also applied to (2,2) Klein space, with the foliation described in Ref.\cite{Casali:2022fro}.} Then
\begin{equation}
\lim_{\rho\rightarrow 0, \,  \tau \rightarrow +\infty}U=x \bar{x} \, ,\quad\lim_{\rho\rightarrow 0, \,  \tau \rightarrow +\infty}  V =(1-x)(1-\bar{x}) \, ,
\end{equation}
where $x$ and $\bar{x}$ are the usual cross ratios on the celestial sphere:
\begin{equation}
x=\frac{z_{12}z_{34}}{z_{13}z_{24}}\, , \quad \bar{x}=\frac{\bar{z}_{12}\bar{z}_{34}}{\bar{z}_{13}\bar{z}_{24}}\ .
\end{equation}
In this limit,
\begin{equation}
\lim_{\rho\rightarrow 0, \,  \tau \rightarrow +\infty}g_\Delta(U,V) =g_\Delta(x,\bar{x}) \, ,
\end{equation}
with $g_\Delta(x,\bar{x})$ given in Eq.(\ref{4dgblexp}).
Moreover, with $d_n=i\lambda_n$, the prefactor of Eq.(\ref{Duffinbls}) becomes
\begin{align}
\lim_{\rho\rightarrow 0, \,  \tau \rightarrow +\infty}&\frac{1}{(x_{12}^2)^{\frac{d_1+d_2}{2}}(x_{34}^2)^{\frac{d_3+d_4}{2}}}\Big( \frac{x_{14}^2}{x_{13}^2}\Big)^{\frac{d_{34}}{2}}\Big( \frac{x_{24}^2}{x_{14}^2}\Big)^{\frac{d_{12}}{2}} \nonumber\\[2mm]
=& \Big(\frac{\rho}{e^{\tau}}\Big)^{i\lambda_1+i\lambda_2+i\lambda_3+i\lambda_4}|z_{12}|^{-i\lambda_1-i\lambda_2}|z_{34}|^{-i\lambda_3-i\lambda_4}\Big|\frac{z_{14}}{z_{13}}\Big|^{i\lambda_3-i\lambda_4}\Big|\frac{z_{24}}{z_{14}} \Big|^{i\lambda_1-i\lambda_2}  \nonumber\\
=&|z_{12}|^{-i\lambda_1-i\lambda_2}|z_{34}|^{-i\lambda_3-i\lambda_4}\Big|\frac{z_{14}}{z_{13}}\Big|^{i\lambda_3-i\lambda_4}\Big|\frac{z_{24}}{z_{14}} \Big|^{i\lambda_1-i\lambda_2} \, ,
\end{align}
where we used $\sum_k i\lambda_k=0$ in the last step.
As a result, the celestial limit of $D=4$ conformal block reads
\be
\lim_{\rho\rightarrow 0, \,  \tau \rightarrow +\infty} W_\Delta(x_n) = |z_{12}|^{-i\lambda_1-i\lambda_2}|z_{34}|^{-i\lambda_3-i\lambda_4}
\Big|\frac{z_{14}}{z_{13}}\Big|^{i\lambda_3-i\lambda_4}\Big|\frac{z_{24}}{z_{14}} \Big|^{i\lambda_1-i\lambda_2} g_\Delta(x,\bar{x}) \, .\label{block11}
\ee

In order to express the celestial amplitude in terms of celestial limits of conformal blocks, we start from Eq.(\ref{DuffinLook}) and apply the following identity derived in Appendix B:
\begin{align}
\Dc_z\;{}_2F_1\left({\alpha,\beta\atop\gamma};z\right)\nonumber
{}_2F_1\left({\alpha,\beta\atop\gamma};\bar z\right)&=-\fc{\alpha\beta}{\gamma}\;\fc{1}{z-\bar z}\lf\{\bar z\; {}_2F_1\left({\alpha,\beta\atop\gamma-1};z\right)
{}_2F_1\left({\alpha+1,\beta+1\atop\gamma+1};\bar z\right)\ri.\\
&\qquad\qquad\qquad\lf.-z\;{}_2F_1\left({\alpha+1,\beta+1\atop\gamma+1};z\right)
{}_2F_1\left({\alpha,\beta\atop\gamma-1};\bar z\right)\ri\}.\label{idapp}
\end{align}
In this way, we obtain
\begin{align}
G(x,\bar{x})&=\delta(\lambda_1+\lambda_2+\lambda_3+\lambda_4)\
\frac{|x|^{i\lambda_1+i\lambda_2}|1-x|^{i\lambda_1-i\lambda_2}}{x(1-x)}
\label{GBlock}\\
&\times\sum_{\Delta=i\lambda_3+i\lambda_4\atop \Delta=4+i\lambda_1+i\lambda_2}^\prime (1-\Delta)\; B\lf(\fc{\Delta-d_{12}}{2},\fc{\Delta+d_{12}}{2}\ri) B\lf(\fc{\Delta+d_{34}}{2},\fc{\Delta-d_{34}}{2}\ri)\;  f_\Delta(z,\bar z)\ ,\nonumber
\end{align}
where
\begin{align}
f_\Delta(z,\bar z)&=\fc{1}{z-\bar z}\; |z|^\Delta\label{BlockD4}\\
&\times\lf\{z\; _2F_1\left( {\h(\Delta-d_{12}),\h(\Delta+d_{43})\atop \Delta}; z\right)
{}_2F_1\left( {\h(\Delta-d_{12})-1,\h(\Delta+d_{43})-1\atop \Delta-2}; \bar z\right)
\ri.\nonumber\\
&\lf.-\bar z\; {}_2F_1\left( {\h(\Delta-d_{12})-1,\h(\Delta+d_{43})-1\atop \Delta-2}; z\right)
{}_2F_1\left( {\h(\Delta-d_{12}),\h(\Delta+d_{43})\atop \Delta}; \bar z\right)\right\}\ .\nonumber
\end{align}
In Eq.(\ref{GBlock}), the prime over the sum accounts for a relative minus sign between the two terms. Next, we use the well-known hypergeometric identity
\be
{}_2F_1\left( {a,b\atop c}; \fc{x}{x-1}\right)=(1-x)^a\ {}_2F_1\left( {a,c-b\atop c}; x\right)\ \label{hyperid}
\ee
to show that
\be
f_\Delta(z,\bar z)= |x-1|^{-d_{12}}\ g_\Delta (x,\bar x)\ ,\label{fgrel}\ee
Finally, we combine all these expressions with Eqs.(\ref{mmms}) and (\ref{partials}), to obtain
\begin{align}
\Big\langle\phi_{\Delta_1,-}^{a_1,-\epsilon}(&z_{1},\bar{z}_{1}
)\,\phi_{\Delta_2,-}^{a_2,-\epsilon}(z_2,\bar{z}_2)
\,\phi_{\Delta_3,+}^{a_3,+\epsilon}
(z_3,\bar{z}_3)\,\phi_{\Delta_4,+}^{a_4,+\epsilon}(z_4,\bar{z}_4)\Big\rangle=
\label{block1}\\
&=\Big(f^{a_1a_2b}f^{a_3a_4b}\,  \frac{1}{z_{12}z_{23}z_{34}z_{41}} + f^{a_1a_3b}f^{a_2a_4b}\,   \frac{1}{z_{13}z_{32}z_{24}z_{41}}\Big)\frac{z_{12}^2}{\bar{z}_{12}^2}\,{\cal S}(z_n,\bz_n)\ ,\nonumber
\end{align}
where
\begin{align}
{\cal S}(z_n&,\bz_n)= |z_{12}|^{-i\lambda_1-i\lambda_2}|z_{34}|^{-i\lambda_3-i\lambda_4}
\Big|\frac{z_{14}}{z_{13}}\Big|^{i\lambda_3-i\lambda_4}\Big|\frac{z_{24}}{z_{14}} \Big|^{i\lambda_1-i\lambda_2} \nonumber\\
&~~~\times\sum^{\prime}_{\Delta=i\lambda_3+i\lambda_4\atop \Delta=4+i\lambda_1+i\lambda_2}\!\!\! (1-\Delta)\; B\lf(\fc{\Delta-d_{12}}{2},\fc{\Delta+d_{12}}{2}\ri) B\lf(\fc{\Delta-d_{34}}{2},\fc{\Delta+d_{34}}{2}\ri)\;  g_\Delta(x,\bar x) \, . \label{fullcorrexp}
\end{align}
By comparing with Eq.(\ref{block11}), we see that ${\cal S}(z_n,\bz_n)$ represents the celestial limit of $D=4$ CFT correlator of four gauge singlet scalars with dimensions $d_n=i\lambda_n$, consisting of
two $D=4$ blocks: one with dimension $\Delta=i\lambda_3+i\lambda_4$ and  one with the ($D=4$) shadow dimension $\Delta=4+i\lambda_1+i\lambda_2=4-i\lambda_3-i\lambda_4$.

In the correlator (\ref{block1}), the first factor on the r.h.s.\ (enclosed in the bracket) represents the group-dependent part of the correlator of four holomorphic (positive helicity) Wess-Zumino-Witten currents \cite{DiF},
\be\label{kac}
\Big\langle J^{a_1}(z_1)J^{a_2}(z_2)J^{a_3}(z_3)J^{a_4}(z_4)\Big\rangle=
k^2\frac{\delta^{a_1a_2}\delta^{a_3a_4}}{z_{12}^2z_{34}^2}-
k\frac{f^{a_1a_2b}f^{ba_3a_4}}{z_{12}z_{23}z_{34}z_{41}}+(2\leftrightarrow 3)
+(2\leftrightarrow 4).
\ee
where $k$ denotes the level of  Ka\v{c}-Moody algebra.
The group-independent part is absent in our celestial amplitude because we did not include the effects of gravitational interactions. As in the heterotic superstring theory, the level $k$ could be determined by computing the relative weight of such contributions.
The subsequent factor $z_{12}^2/\bar{z}_{12}^2$, present in Eq.(\ref{block1}), ``flips'' the helicity of gluons number 1 and 2 from $+$ to $-$. Both factors do not depend on $\lambda_n$. The $\lambda_n$-dependence is contained entirely in the scalar factor ${\cal S}(z_n,\bz_n)$. This correlator depends neither on helicities nor on gauge charges of external gluons.
Besides Eq.(\ref{fullcorrexp}), the scalar part of the correlator can be also represented in a Coulomb gas form as
\begin{align}
{\cal S}(z_n,\bz_n)=&~|z_{12}|^{-2i\lambda_1-2i\lambda_2}|z_{13}|^{-2i\lambda_1-2i\lambda_3} |z_{14}|^{-2i\lambda_4} |z_{23}|^{2i\lambda_1}\nonumber \\ &\times (2+i\lambda_1+i\lambda_2)B(2+i\lambda_1,1+i\lambda_2)\label{liou}
B(i\lambda_3,-1+i\lambda_4)\\ &\times\Dc_z \Big(|1-x|^{-2(i\lambda_1+1)}\;\int d^2w\ |w|^{-2i\lambda_4+2}\ |w-1|^{-2i\lambda_2-2}\ |w-x|^{-2i\lambda_3} \Big)\, \nonumber
\end{align}
where $\Dc_z$ is given in Eq.(\ref{opdczx}). This suggest that the CCFT operators associated to gluons factorize into the current parts and certain operators similar to those encountered in Coulomb gas models and in the infinite central charge limit of Liouville theory \cite{Dorn:1994xn}. In string theory, the integrals of the form (\ref{liou}), with the kinematic invariants (instead of dimensions
$i\lambda_n$) in the exponents, are also known as the Koba-Nielsen integrals.
\section{$D=2$ conformal block decomposition of $D=4$ blocks}
As concluded in the previous section, we expect $D=4$ conformal blocks (\ref{fullcorrexp}) to describe an autonomous sector of CCFT.  This motivates us to perform $D=2$ conformal decomposition of this correlator. The $D=2$ chiral weights of external scalars are $h_k=\bh_k=i\lambda_k/2$. We consider the conformal block decomposition of\footnote{Here, we ignore the $\delta(\lambda_1+\lambda_2+\lambda_3+\lambda_4)$ prefactor.}
\be
G_{\cal S}(x,\bar{x}) ~=~ \lim_{z_1,\bar{z}_1\rightarrow \infty} z_1^{i\lambda_1} \bar{z}_1^{i\lambda_1}{\cal S}(z_1; z_2=1; z_3=x; z_4=0) ~=~x(1-x)\, G(x,\bar{x})\ .
\ee

In Ref.\cite{Fan:2022vbz}, we performed the conformal block decomposition of
the four-gluon correlator $G(x,\bar{x})$ by using the power series solution of BG equations. The $s$-channel ($x\approx 0$) conformal blocks associated to primary fields with chiral weights $(h,\bar{h})$ were denoted by $K_{34}^{21}\left[h,\bar{h}\right](x,\bx)$ and their explicit form was given in Ref.\cite{Fan:2022vbz}. The conformal block decomposition of $G_{\cal S}(x,\bar{x})$ can be performed in a similar way, with the following result:
\begin{align}
	G_{\cal S}(x,\bar{x}) 	= \sum_{n=0}^{\infty}&\left\{ a_{n} \, K_{34}^{21}\left[n+\frac{i \lambda_{3}+i \lambda_{4}}{2}, n+ \frac{i \lambda_{3}+i\lambda_{4}}{2}\right] (x,\bx)\right. \\  & +\left. b_{n}\, K_{34}^{21}\left[n+2-\frac{i \lambda_{3}+i \lambda_{4}}{2}, n+2 -\frac{i\lambda_{3}+i\lambda_{4}}{2}\right](x,\bx)\right\} \, .
\end{align}
with the coefficients
\begin{align}
	a_{n}&= B(n-i \lambda_{1}, n-i \lambda_{2} ) B(n+i \lambda_{3},  n +i \lambda_{4}) \left(1-2 n+i \lambda_{1}+i \lambda_{2}\right) ,\\[1mm]
	b_{n}&= -B(n+2-i \lambda_{3},  n+2-i \lambda_{4} ) B(n+2+i \lambda_{1}, n+2+i \lambda_{2} ) \left(-3-2 n+i \lambda_{3}+i \lambda_{4}\right).\nonumber
\end{align}
There are two sets of conformal blocks propagating in the $s$-channel: one
with the chiral weights
\be h_{n}=\bar{h}_{n}= n+ \frac{i \lambda_{3}+i\lambda_{4}}{2}\ee
and one with
\be h_{n}={\bar{h}}_{n}=n+2-\frac{i \lambda_{3}+i \lambda_{4}}{2}\ .\ee
Note that all blocks have zero spin. We stress that they are gauge singlets because the group-dependence is contained in the separate, Wess-Zumino-Witten part of the four-gluon correlator.\footnote{The block decomposition presented here can be viewed as a dimensional reduction of conformal blocks from $D$ to $D{-}2$ dimensions, see Ref.\cite{Hogervorst:2016hal}. For the reverse case, see Ref.\cite{Kaviraj:2019tbg}, where $(D{-}2)$-dimensional blocks are written as finite linear combinations of $D$-dimensional blocks.}
\section{Conclusions}
In this work, we studied the properties of four-gluon celestial amplitudes constructed in Ref.\cite{Fan:2022vbz}. We focused on the integral representation describing four-gluon scattering in the presence of a dilaton background.
We expressed it in terms of Appell's $F_4$ hypergeometric function and showed that it agrees with the series expansion obtained by solving Banejree-Ghosh equations.

Our  main result is that the four-gluon celestial amplitude factorizes into the ``current'' part and the ``scalar'' part, as written in Eq.(\ref{block1}). The current factor is given by the group-dependent part of the Wess-Zumino-Witten correlator of four holomorphic currents with a non-vanishing level of Ka\v{c}-Moody algebra. The scalar factor can be expressed in terms of a complex integral of the Koba-Nielsen form, similar to the integrals describing four-point correlators in Coulomb gas models and, more generally, in the infinite central charge limit of Liouville theory. The connection to Liouville theory is not precise at this point, but we believe that once it is put on solid grounds, it will establish a close connection between CCFT and Liouville theory. CCFT is similar to the heterotic superstring theory, which also contains an ``internal'' holomorphic WZW current sector, but the ``spacetime''
sector of heterotic strings is replaced in CCFT by Liouville theory.

A very interesting feature of the scalar part of the correlator is that it can be obtained by a dimensional reduction of a single $D=4$ conformal block and the shadow block from Minkowski space to the celestial sphere. This is certainly related to $D=4$ conformal symmetry of Yang-Mills theory at the tree level, but the fact that a ``minimal''  number of one block and its shadow appear in the amplitude  must have a deeper explanation. Understanding the role of conformal symmetry and its breaking by quantum loops and by nonperturbative effects in CCFT will  help in studying Yang-Mills dynamics from a novel angle.\\[1ex]\noindent {\bf Acknowledgments}\\[2mm]
This material is based in part upon work supported by the National Science Foundation
under Grant Number PHY--1913328.
Any opinions, findings, and conclusions or recommendations
expressed in this material are those of the authors and do not necessarily
reflect the views of the National Science Foundation.
Wei Fan is supported in part by the National Natural Science Foundation of China under Grant No.\ 12105121.

\appendix

\section{Intermediate steps from Eq.(\ref{eq:G1_IntRep}) to Eq.\req{DuffinLook}}
By using the integral representation of the hypergeometric function $_2F_1$,
$$_2F_1\left( {a,b\atop c}; x \right)=B(b,c-b)^{-1}\ \int_0^1d\eta\; \eta^{b-1}\;(1-\eta)^{c-b-1}\; (1-\eta x)^{-a}\ ,$$
we can express the integral (\ref{eq:G1_IntRep}) as the following double integral
\begin{align}
&G (x,\bar{x})=\delta(\lambda_1+\lambda_2+\lambda_3+\lambda_4)\ \frac{1}{x(1-x)}\\
&{\textstyle \times\Bigg\{ (1+i\lambda_1)
\int_0^1 d\xi\; \xi^{i\lambda_3-1}\; (1-\xi)^{i\lambda_1+i\lambda_4-1} \int_0^1d\eta\; \eta^{-i\lambda_2-1}\; (1-\eta)^{-2-i\lambda_1}\; \lf(1+\fc{\xi u_2}{1-\xi}-\xi\eta u_1\ri)^{i\lambda_1}\nonumber}\\
&{\textstyle +(1+i\lambda_2)\;u_1^{2+i\lambda_1+i\lambda_2}\;
\int_0^1 d\xi\; \xi^{1-i\lambda_4}\; (1-\xi)^{-1-i\lambda_3-i\lambda_2} \int_0^1d\eta\; \eta^{1+i\lambda_1}\; (1-\eta)^{i\lambda_2}\; \lf(1+\fc{\xi u_2}{1-\xi}-\xi\eta u_1\ri)^{-2-i\lambda_2}\Bigg\},\nonumber }
\end{align}
with:
\be
u_1=|x|^2\ \ \ ,\ \ \ u_2=|1-x|^2\ .
\ee
After introducing the new integration variables,
\be
\xi=\fc{1}{1+\beta}\ \ \ ,\ \ \ \eta=\frac{1+\beta}{\gamma}\ ,
\ee
with $d\xi d\eta=\tfrac{d\beta d\gamma}{(1+\beta)\gamma^2}$, we obtain
\begin{align}
G (x,\bar{x})&=\delta(\lambda_1+\lambda_2+\lambda_3+\lambda_4)\ \frac{1}{x(1-x)}\label{PreDuff}\\
 &\times\Bigg\{
(1+i\lambda_1)\int_0^\infty \fc{d\beta}{\beta}\int_{1+\beta}^\infty\fc{d\gamma}{\gamma}\  \fc{\beta^{i\lambda_4}\;\gamma^{2+i\lambda_2}}{(\beta\gamma+\gamma u_2-\beta u_1)^{-i\lambda_1}\;(\gamma-\beta-1)^{2+i\lambda_1}}\nonumber\\
&+(1+i\lambda_2)\lf(\fc{u}{v}\ri)^{2+i\lambda_1+i\lambda_2}\int_0^\infty \fc{d\beta}{\beta}\int_{1+\beta}^\infty\fc{d\gamma}{\gamma}\  \fc{\beta^{2-i\lambda_3}\;\gamma^{-i\lambda_1}}{(\beta\gamma+\gamma u_2-\beta u_1)^{2+i\lambda_2}\;(\gamma-\beta-1)^{-i\lambda_2}}\Bigg\},\nonumber
\end{align}
with:
\begin{align}
v&=\fc{1}{u_2}=\fc{1}{|1-x|^2}=|z-1|^2\ ,\\
u&=\fc{u_1}{u_2}=\lf|\fc{x}{x-1}\ri|^2=|z|^2\ .\label{variables}
\end{align}
Furthermore, \req{PreDuff} can  be expressed as:
\begin{align}
G& (x,\bar{x})=-\delta(\lambda_1+\lambda_2+\lambda_3+\lambda_4)\ \frac{v^{-i\lambda_1}}{x(1-x)}\ \label{Duffinlook}\\
&\times\fc{\p}{\p v}\ \Bigg\{
\int_0^\infty \fc{d\beta}{\beta}\int_{1+\beta}^\infty
\fc{d\gamma}{\gamma}\; \fc{\beta^{i\lambda_4-1}\;\gamma^{1+i\lambda_2}}{(\gamma+\beta\gamma v-\beta u)^{-1-i\lambda_1}\;(\gamma-\beta-1)^{2+i\lambda_1}}\nonumber\\
&\qquad\qquad-u^{2+i\lambda_1+i\lambda_2}\int_0^\infty \fc{d\beta}{\beta}\int_{1+\beta}^\infty
\fc{d\gamma}{\gamma}\; \fc{\beta^{1-i\lambda_3}\;\gamma^{-1-i\lambda_1}}{(\gamma+\beta\gamma v-\beta u)^{1+i\lambda_2}\;(\gamma-\beta-1)^{-i\lambda_2}}\Bigg\}\ .\nonumber
\end{align}
Now we can borrow the following integral \cite{Simmons-Duffin:2012juh}
\begin{align}
\int_0^\infty \fc{d\beta}{\beta}\int_{1+\beta}^\infty
\fc{d\gamma}{\gamma}&\; \fc{\beta^{b-1}\;\gamma^{e-1}}{(\gamma+v\beta\gamma-u\beta)^{1-f}\;(\gamma-\beta-1)^f}\nonumber\\
&=B(-1+b,4-b-e-f)\; B(2-e,1-f)\ \\
&\times {}_2F_1\left( {-1+b, 1-f \atop 3-e-f}; z\right)\; _2F_1\left( {-1+b, 1-f \atop 3-e-f}; \bar z\right)\ ,\nonumber
\end{align}
to cast (\ref{Duffinlook}) into Eq.\req{DuffinLook}.

\section{Hypergeometric function relations}

In this appendix we detail the steps from the integral representation Eq.\req{DuffinLook} to Eq.\req{ExplicitF4}.

We first notice the duplication  formula \cite{Slater}
\be\label{duplication}
{}_2F_1\left({\alpha,\beta\atop\gamma};z\right)
{}_2F_1\left({\alpha,\beta\atop\gamma};\bar z\right)=\cG\lf[\alpha,\beta,\gamma,\gamma;u,1-v\ri]\ ,
\ee
with the variables related as in Eq.\req{variables}, and
\begin{align}
\cG\lf[\alpha,\beta,\gamma,\delta;u,1-v\ri]&=\fc{\Gamma(\delta)\Gamma(\delta-\alpha-\beta)}{\Gamma(\delta-\alpha)\Gamma(\delta-\beta)}\ F_4\lf[\alpha,\beta,\gamma,\alpha+\beta+1-\delta;u,v\ri]\nonumber\\
&+\fc{\Gamma(\delta)\Gamma(\alpha+\beta-\delta)}{\Gamma(\alpha)\Gamma(\beta)}\; v^{\delta-\alpha-\beta}\;F_4\lf[\delta-\alpha,\delta-\beta,\gamma,\delta-\alpha-\beta+1;u,v\ri]\nonumber\\
&=\sum_{m,n=0}^\infty\fc{u^m(1-v)^n}{m!n!}\;\fc{(\delta-\alpha)_m(\delta-\beta)_m}{(\gamma)_m}\fc{(\alpha)_{m+n}(\beta)_{m+n}}{(\delta)_{2m+n}}\ .\label{Gfunction}
\end{align}
The function \req{Gfunction} enjoys the transformation property \cite{Dolan:2000uw}:
\be
\cG\lf[\alpha,\beta,\gamma,\delta;u,1-v\ri]=v^{-\alpha}\ \cG\lf[\alpha,\delta-\beta,\gamma,\delta;\fc{u}{v},1-\fc{1}{v}\ri]\ .\label{symmGtrans}
\ee
Next, we consider the relation
\be
\p_v\;\cG\lf[\alpha,\beta,\gamma,\delta;u,1-v\ri]=-\fc{\alpha\beta}{\delta}\ \cG\lf[\alpha+1,\beta+1,\gamma,\delta+1;u,1-v\ri]\ ,
\ee
following from
\be
\p_v F_4\lf[\alpha,\beta,\gamma,\delta;u,v\ri]=\fc{\alpha\beta}{\delta}\ F_4\lf[1+\alpha,1+\beta,\gamma,1+\delta;u,v\ri]\ ,
\ee
which together with \req{duplication} gives rise to:
\begin{align}
\Dc_z\;{}_2F_1\left({\alpha,\beta\atop\gamma};z\right)
{}_2F_1\left({\alpha,\beta\atop\gamma};\bar z\right)&=\fc{\alpha\beta}{\gamma}\ \cG\lf[\alpha+1,\beta+1,\gamma,\gamma+1;u,1-v\ri]\label{DuplicationD}\\
&=-\fc{\alpha\beta}{\gamma}\;\fc{1}{z-\bar z}\lf\{\bar z\; {}_2F_1\left({\alpha,\beta\atop\gamma-1};z\right)
{}_2F_1\left({\alpha+1,\beta+1\atop\gamma+1};\bar z\right)\ri.\nonumber\\
&\qquad\qquad\qquad\lf.-z\;{}_2F_1\left({\alpha+1,\beta+1\atop\gamma+1};z\right)
{}_2F_1\left({\alpha,\beta\atop\gamma-1};\bar z\right)\ri\}\nonumber
\end{align}
After inserting Eqs.(\ref{DuplicationD}) and (\ref{Gfunction}) into Eq.\req{DuffinLook}, we obtain Eq.\req{ExplicitF4}.

\end{document}